\documentclass{ws-p8-50x6-00}

\def\be{\begin{equation}}
\def\ee{\end{equation}}
\def\bea{\begin{eqnarray}}
\def\eea{\end{eqnarray}}

\begin{document}

\title{(Some of) recent $\gamma\gamma$ measurements from LEP}

\author{\v{S}.Todorova-Nov\'{a}}

\address{CERN, CH-1211 Geneva, Switzerland\\E-mail:todorovova@cern.ch}

\maketitle
\abstracts{Inclusive measurement of
$J/\Psi$ (DELPHI), search for $\eta_b$ (ALEPH) and measurement
 of the cross-section  of the double-tagged $\gamma\gamma$ events
(ALEPH) are briefly overviewed.  }

\section{Measurement of inclusive  J/$\Psi$ production in 
 $\gamma\gamma$ events (DELPHI)} 

   The measurement of J/$\Psi$ production in $\gamma\gamma$
   interactions provides an insight into heavy quark
   production and also a possibility to study the gluon parton density
   of a resolved photon.

   Selection of the signal is done in two steps. First, preselection of hadronic
  $\gamma\gamma$ events is performed, defining the ``visible'' region
  (visible inv.mass $W_{vis}$ below 35 GeV/c$^2$,
 charged multiplicity $ 4 \leq N_{ch} \leq 30 $, transverse energy
  above 3 GeV, cuts ensuring high trigger efficiency).
  The preselected sample contains about $1.2\%$ of $Z^0/\gamma^*$
  background. Please note that requirement of at least 4 reconstructed
  tracks suppress the $J/\Psi+\gamma$ final state.
   
\begin{figure}[bth]
\mbox{\epsfig{file=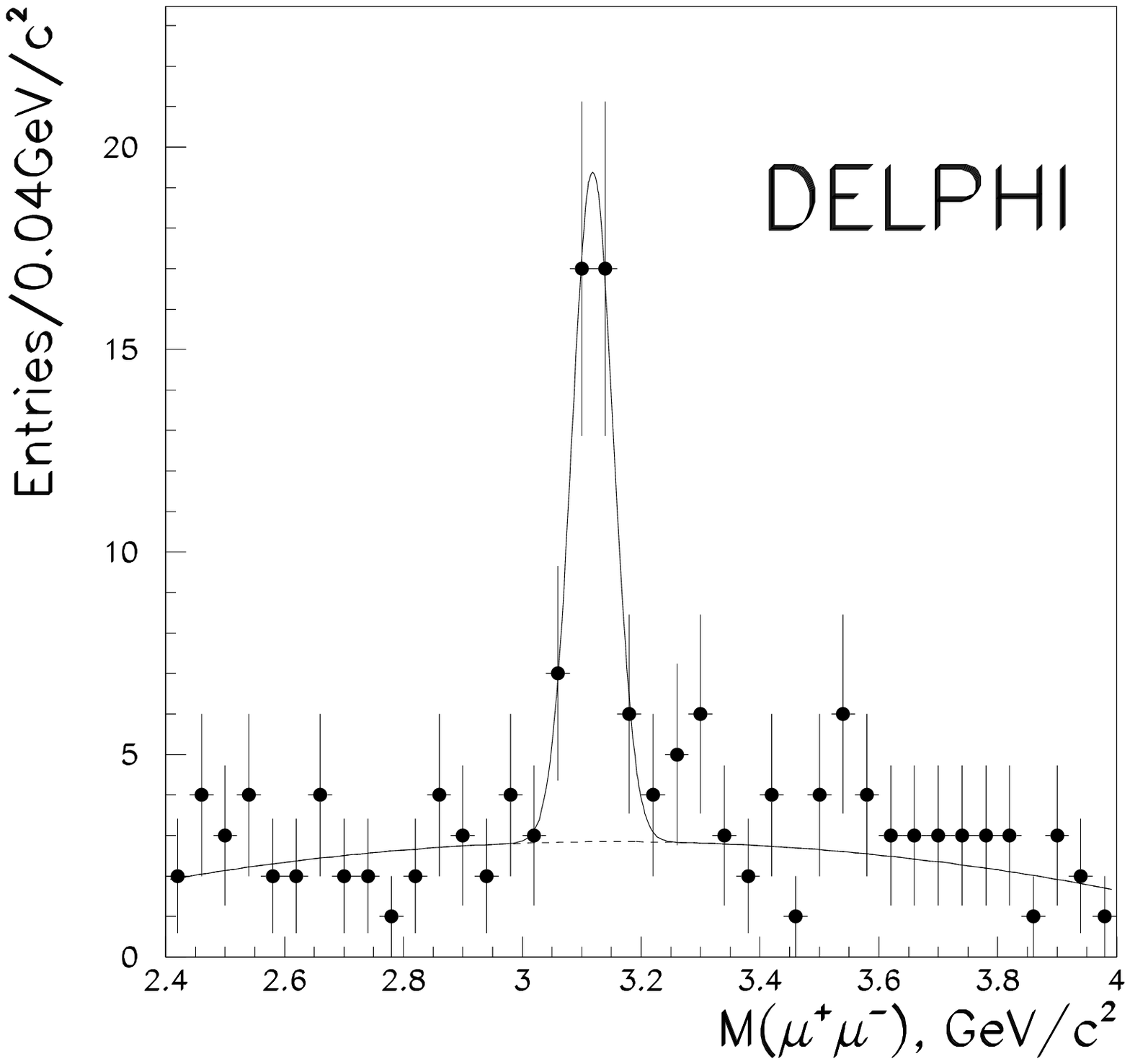,height=6cm}
      \epsfig{file=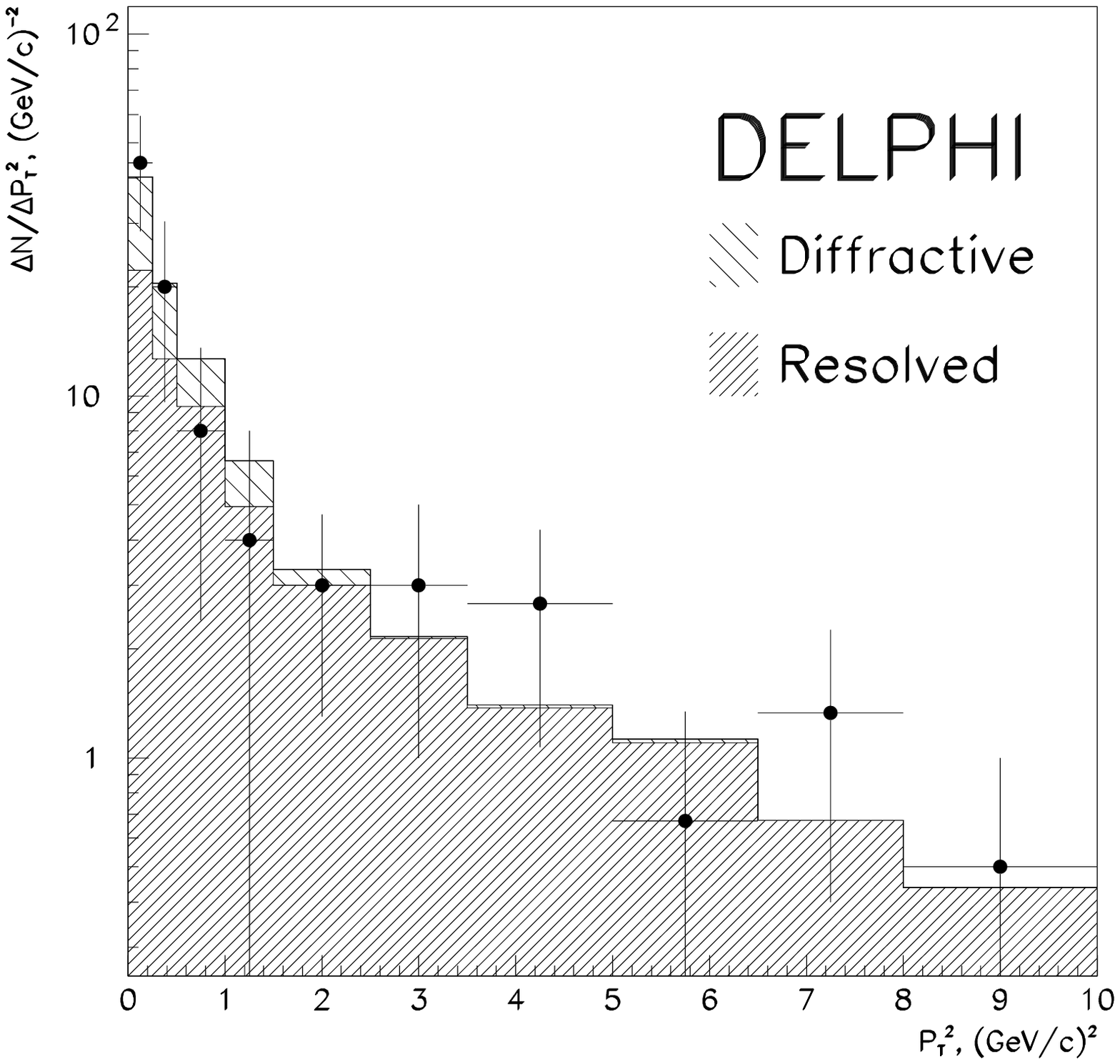,height=6.cm}}
\caption{\small \sl Left: Reconstructed mass spectrum of identified muon pairs.
Right: The best fit of the transverse momentum of the reconstructed $J/\Psi$
with a mixture of resolved and diffractive events simulated with
Pythia 6.1.}
 \label{fig:jpsi}
\end{figure}

   Next, the J/$\Psi$ signal is obtained requiring a pair of
 identified muons. The mass spectrum of the muon pair in the region of interest
 is shown in Fig.\ref{fig:jpsi}a, with a clear J/$\Psi$ signal
 ($ 36 \pm 7 $ events)
 on top
 of combinatorial $\gamma\gamma$ background.
 The fit of the mass spectrum gives 
 $ M = 3.119 \pm 0.008 $ GeV/c$^2$, $ \Gamma(obs) = 0.035 \pm
0.007$ GeV.

  In order to learn more about the production mechanism (the absolute
 predictions of Pythia 6.156~\cite{pythia} fall well below ($\sim 20\%$) the observed
 production rate), the relative fraction of resolved and diffractive
 processes was fitted to the shape of the transverse
 momentum squared of the reconstructed $J/\Psi$. For this study,
 the shapes of Pythia distributions, with full detector simulation
 and event selection, were used. Fig.\ref{fig:jpsi}b shows the best fit   
 \begin{eqnarray*}
 \frac{d{\it N}}{d{\it p_t^2}} & = &  f \cdot \frac{d{\it N}}{d{\it
 p_t^2}}\mid_{Resolved} 
    +  (1-f) \cdot \frac{d{\it N}}{d{\it p_t^2}}\mid_{Diffractive},  \\
  & & \\
  f & = & (74.0 \pm 22.0) \% , 
 \end{eqnarray*}
   suggesting the dominant contribution of resolved processes, in
 agreement with the expectation (the fit is obviously driven by the tail
 of the transverse momentum distribution). Within the statistics
 available, the renormalized mixture of Pythia processes describes
data reasonably well (including multiplicity and invariant mass of the
 recoiling system), and the simulated data can be used to correct the
 visible cross-section.

 The production rate per preselected event, using  
  $Br(J/\Psi\rightarrow\mu^+\mu^-$) = (5.88$\pm$0.1)\%), and the visible cross-section are estimated at 
\begin{description}
\item $ <n> = N(J/\Psi)\cdot(N_{ev}\cdot Br \cdot \epsilon_{vis})^{-1} =
(6.7 \pm 1.3) \mathrm{x} 10^{-3} $ 
\item $  \sigma_{vis}(\gamma\gamma\rightarrow J/\Psi+X) =
  N(J/\Psi)\cdot(Br \cdot \epsilon_{vis} \cdot \mathcal{L} \mathrm)^{-1} =  2.98 \pm 0.58 $ pb. 
\end{description}

  Work is going on direct comparison with NLO calculations
  (in the experimentally accessible part of the phase space).

\section{Search for $\eta_b$  in 
 $\gamma\gamma$ events (ALEPH) } 

   The search for $\eta_b$ (the ground $b\bar{b}$ state) in
   $\gamma\gamma$ interactions is motivated by the possibility
   of the exclusive production of this resonance ($J^{PC}=0^{+-}$).

    The mass of the $\eta_b$ should lie just below the $Y$
    vector meson, and various calculations (lattice, potential model,
    ...) predict a range of 9.36-9.42 GeV. The production rate ($\sim$
    0.222 pb at $\sqrt{s}=197$ GeV) was
    estimated using the Coulomb potential approach and the measured
    $\Gamma_{\gamma\gamma}(\eta_c)$. 
     
    The data used for the analysis were collected by ALEPH at the mean 
    energy of  $\sqrt{s}=197$ GeV and correspond to the integrated
     luminosity of 700 pb$^{-1}$. The
    search for $\eta_b$ was performed in the 4 and 6 charged particles
    mode. Events with a neutral object, or identified muon/electron(s)
    were rejected. The total transverse momentum of charged tracks
    below 0.25 GeV was required, and further cuts on the thrust
    ($<0.95$) and polar angle of the thrust axis ($>45^o$) were
    added to suppress $\gamma\gamma$ continuum and
    $\tau\tau$ background (the latter was then reduced to the negligible
    fraction by direct cuts on the net charge and mass of particles
    in both hemispheres).  
  
      The selection and reconstruction efficiency were studied with
    help of simulated PHOT02\cite{phot02} events, and found to be 15.7\% and 10.1\%
    in the 4 and 6 charged particles modes, respectively. The
    background, dominated by the $\gamma\gamma $ continuum, was
    estimated using the data rather than MC, and found to be 0.3$\pm$ 0.3
    (0.8$\pm$0.4) events in the 4(6) charged particle mode.

\vspace{-0.5cm}

\begin{figure}[bht]
\begin{center}
\mbox{\epsfig{file=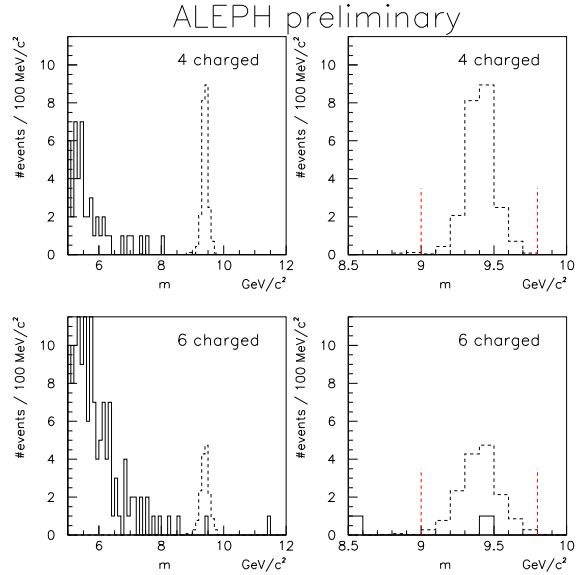,width=9cm}}
\caption{\small \sl Invariant mass spectrum of selected events.}
\label{fig:etab}
\end{center}    
\end{figure}
       Invariant spectra of selected events are shown in
 Fig.~\ref{fig:etab}. In the
     signal region (9.0-9.8 GeV), 0 (resp.1) candidates are left
     after all cuts in the 4 (resp. 6) particle mode (the dashed line
     indicates the expected signal assuming the corresponding
     branching ratio is 100\%). The selected candidate is compatible
     with background expectations.

       The measurement allows to set an upper limit on the
     partial widths and
     branching ratios of $\eta_b$ into decay channels under study:      
     Br($\eta_b \rightarrow $ 4 charged particles) $<$ 17 \% ,
      Br($\eta_b \rightarrow $ 6 charged particles) $<$ 38 \%  at the 95
     \% confidence limit (using the estimated  
     $\Gamma_{\gamma\gamma}$($\eta_b$) = 416 eV)\cite{etab}.

\section{Double tagged hadronic cross section (ALEPH)}

   The double-tagged $\gamma\gamma$ events were studied by ALEPH 
  using the LEP2 data (640 pb$^{-1}$ at c.m.s. energies between
  189-208 GeV). The main selection criteria consist in detection
  of both scattered electrons - carrying at least 0.3 of the beam
  energy - in the polar angle range 35 mrad $<
  \Theta <$ 55 mrad. Additional cuts are imposed on the hadronic
  final state: visible mass above 3 GeV, at least 3 reconstructed 
  charged tracks. Cuts on the minimal total energy (0.7 E$_{cms}$)
  and maximal opening angle of leptons (179.5$^o$) help to suppress the
  $\tau\bar{\tau}$ background. The selected sample consists of 891
  events  (background expectation amounts to 206.1 events).

 \vspace{-0.5cm}

\begin{figure}[bth]
\begin{center}
\mbox{\epsfig{file=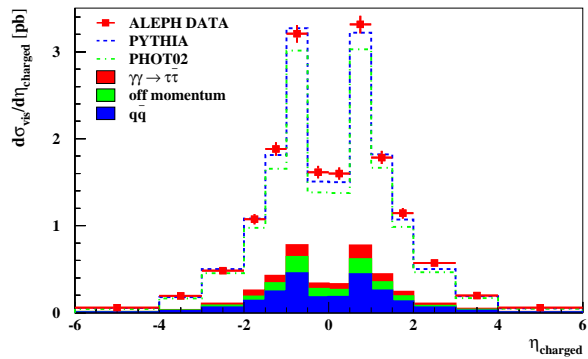,width=8.5cm}}
\caption{\small \sl Pseudorapidity disribution at the detector level.
 MC predictions are scaled to match the observed number of events.}
\label{fig:a_eta}
\end{center}
\end{figure}

      The measured spectra are generally in a good agreement with
  simulation after renormalization to the observed number
  of events. For example, Fig.\ref{fig:a_eta} shows the pseudorapidity
  distribution observed in data versus MC predictions. Pythia 6.151
  was rescaled by 30 \% and PHOT02 by 12 \% for this comparison.

\begin{figure}[h]
\begin{center}
\mbox{\epsfig{file=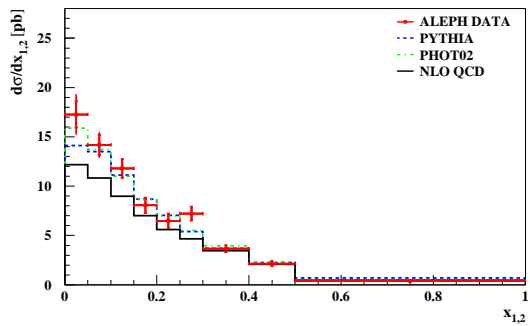,width=7cm}}
\caption{\small \sl Bj\"{o}rken x distribution of the corrected data,
 compared to the 
 MC predictions (scaled to match the observed number of events) and
 NLO QCD calculations.}
\label{fig:x}
\end{center}
\end{figure}

     After background subtraction and application of bin corrections
   for the detector acceptance, the differential cross-section was
   compared both with MC and NLO QCD calculations\cite{nlo}. The NLO
   calculations tend to underestimate the data, see Fig.~\ref{fig:x}.

\begin{figure}[bth]
\begin{center}
\mbox{\epsfig{file=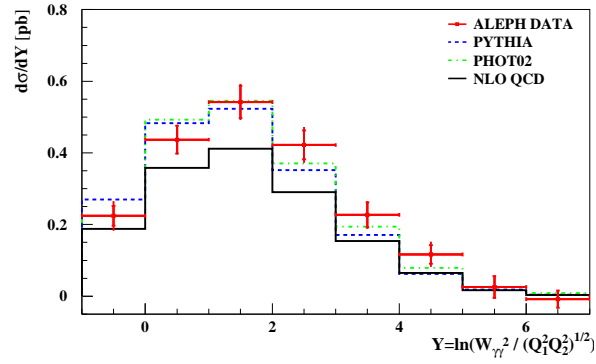,width=8.5cm}}
\caption{\small \sl Y distribution of the corrected data,
 compared to the 
 MC predictions (scaled to match the observed number of events) and
 NLO QCD calculations.}
\label{fig:y}
\end{center}
\end{figure}

\begin{figure}[bth]
\begin{center}
\mbox{\epsfig{file=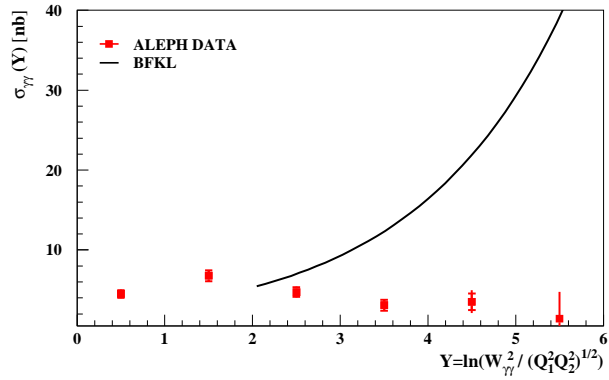,width=8cm}}
\caption{\small \sl Comparison of the unfolded  differential $\gamma\gamma$
 cross-section (as a function of the relative inv.mass 
  of the $\gamma\gamma$
 system, $\sim Y$ variable) with the LL BFKL calculations.}
\label{fig:bfkl}
\end{center}
\end{figure}

    A study of BFKL was done using the approximate relation
   $Y = ln(s_{ee}/s_{\gamma\gamma}) \approx
    ln(W_{\gamma\gamma}^2/\sqrt{Q_1^2 Q_2^2}) $, valid for $
   W_{\gamma\gamma}^2 \gg Q^2 $ (the differential cross section in this
   variable is plotted in Fig.~\ref{fig:y}). For comparison with BFKL
   calculations, additional cut log$(Q^2_1/Q^2_2)<1.0$ was added, ensuring
   both photons have similar virtuality, and
   the $\sigma_{\gamma\gamma}$ was extracted from
   the double tagged cross-section using photon flux calculated by
   GALUGA\cite{galuga}. The comparison with LL BFKL\cite{bfkl} is shown in
   Fig.~\ref{fig:bfkl}. Data do not show the enhancement with $Y$,
   which is - at least partly - understood (next-to-leading-log
   BFKL resummation brings large negative contribution; kinematical
   effects - like energy conservation - are neglected). Similar results
   were reported also by L3\cite{l3} and OPAL\cite{opal}.

\begin{thebibliography}{99}

\bibitem{pythia}T.Sj\"{o}strand et al., {\it Comput. Phys. Commun.} 135:238, 2001.
\bibitem{etab}ALEPH Coll., ALEPH 2001-037 CONF 2001-025.
\bibitem{phot02}A.Finch, PHOT02 Monte Carlo Generator.
\bibitem{nlo}Cacciari et al., {\it JHEP}, 02:029, 2001.
\bibitem{galuga}G.Schuller, {\it Comput. Phys. Commun.}, 108:279, 1998.
\bibitem{bfkl}Bartels et al., {\it Phys.Lett.}, B429:56-65, 2000.
\bibitem{l3}L3 Coll., {\it Phys.Lett.}, B453:333, 1999.
\bibitem{opal}OPAL Coll., OPAL Physics Note PN456, 2000.
\end{thebibliography}
\end{document}